\documentclass[journal]{IEEEtran}

\usepackage{hyperref}
\hypersetup{
	colorlinks = true, 
	urlcolor  = cyan, 
	linkcolor = blue, 
	citecolor = red 
}
\usepackage{url}
\usepackage{amsmath}
\usepackage{amsfonts}
\usepackage{amssymb}
\usepackage{graphicx}
\usepackage{amsthm}
\usepackage{mathbbol}
\usepackage{dsfont}
\usepackage{color}
\usepackage{makecell}
\usepackage{booktabs}
\usepackage{algorithm}
\usepackage{algorithmicx}
\usepackage{algpseudocode}

\begin{document}

\title{Unsupervised Congestion Status Identification Using LMP Data}

%
%
%

\author{Kedi~Zheng,~\IEEEmembership{Student~Member,~IEEE,}
	Qixin~Chen,~\IEEEmembership{Senior~Member,~IEEE,}
	Yi~Wang,~\IEEEmembership{Member,~IEEE,} 
	Chongqing~Kang,~\IEEEmembership{Fellow,~IEEE,}
	and Le~Xie~\IEEEmembership{Senior~Member,~IEEE}
\thanks{Manuscript received October 7, 2019; revised February 28, 2020, and June 15, 2020; accepted July 19, 2020. This work was support by Major Smart Grid Joint Project of National Natural Science Foundation of China and State Grid (No. U1766212). Paper no. TSG-01488-2019. (\textit{Corresponding author: Qixin Chen}.) 

K. Zheng, Q. Chen, and C. Kang are with the State Key Lab of Power Systems, Department of Electrical Engineering, Tsinghua University (email: zkd17@mails.tsinghua.edu.cn; qxchen@tsinghua.edu.cn; cqkang@tsinghua.edu.cn).

Y. Wang is with Power Systems Laboratory, Department of Information Technology and Electrical Engineering, ETH Z\"urich, Z\"urich 8092, Switzerland (email: yiwang@eeh.ee.ethz.ch).

L. Xie is with Department of Electrical \& Computer Engineering, Texas A\&M University (email: le.xie@tamu.edu).
}
\thanks{Digital Object Identifier \href{https://doi.org/10.1109/TSG.2020.3011266}{10.1109/TSG.2020.3011266}}
}

%
%

\markboth{IEEE TRANSACTIONS ON SMART GRID, MANUSCRIPT}%
{Shell \MakeLowercase{\textit{et al.}}: Bare Demo of IEEEtran.cls for IEEE Journals}
%


\maketitle

\IEEEpubidadjcol

\IEEEpubid{\begin{minipage}{\textwidth}\ \\[12pt] \centering
		© 2021 IEEE.  Personal use of this material is permitted.  Permission from IEEE must be obtained for all other uses, in any current or future media, including reprinting/republishing this material for advertising or promotional purposes, creating new collective works, for resale or redistribution to servers or lists, or reuse of any copyrighted component of this work in other works.
\end{minipage}}

\begin{abstract}
Having a better understanding of how locational marginal prices (LMPs) change helps in price forecasting and market strategy making. This paper investigates the fundamental distribution of the congestion part of LMPs in high-dimensional Euclidean space using an unsupervised approach. LMP models based on the lossless and lossy DC optimal power flow (DC-OPF) are analyzed to show the overlapping subspace property of the LMP data. The congestion part of LMPs is spanned by certain row vectors of the power transfer distribution factor (PTDF) matrix, and the subspace attributes of an LMP vector uniquely are found to reflect the instantaneous congestion status of all the transmission lines. The proposed method searches for the basis vectors that span the subspaces of congestion LMP data in hierarchical ways. In the bottom-up search, the data belonging to 1-dimensional subspaces are detected, and other data are projected on the orthogonal subspaces. This procedure is repeated until all the basis vectors are found or the basis gap appears. Top-down searching is used to address the basis gap by hyperplane detection with outliers. Once all the basis vectors are detected, the congestion status can be identified. Numerical experiments based on the IEEE 30-bus system, IEEE 118-bus system, Illinois 200-bus system, and Southwest Power Pool are conducted to show the performance of the proposed method.

\end{abstract}


%
\IEEEpeerreviewmaketitle

\section*{Nomenclature}

\IEEEpubidadjcol

\addcontentsline{toc}{section}{Nomenclature}

\noindent \textit{Sets and Indices}
\begin{IEEEdescription}
	\item[$\mathcal{N}$] Set of all nodes in an electricity market.
	\item[$\mathcal{T}$] Set of time intervals.
	\item[$\mathcal{B}$] Set of basis vectors.
	\item[$t$] Time interval index. 
	\item[$i$] Node index.
	\item[$j$] Line index.
\end{IEEEdescription}

\noindent \textit{Variables, Parameters and Functions}
\begin{IEEEdescription}[\IEEEusemathlabelsep\IEEEsetlabelwidth{$ \boldsymbol{P}_G, \boldsymbol{P}_D $}]
	\item[$n_b, n_l $] Number of buses and lines, respectively.
	\item[$ \boldsymbol{P}_G, \boldsymbol{P}_D $] Generated and consumed active power vectors, respectively.
	\item[$ P_{Gi}, P_{Di} $] Generated and consumed power at node $ i $, respectively.
	\item[$ \boldsymbol{T} $] Power transfer distribution factor (PTDF) matrix.
	\item[$ \boldsymbol{f}^{\max} $] Transmission capacity vector.
	\item[$ \boldsymbol{P}_G^{\min/\max} $] Generation output limit. 
	\item[$ \boldsymbol{1} $] All-ones vector.
	\item[$ \lambda $] Shadow price for the balance constraint.
	\item[$ \boldsymbol{\mu}^{+/-}, \boldsymbol{\mu} $] Shadow prices for transmission constraints. $ \boldsymbol{\mu} = \boldsymbol{\mu}^{-}-\boldsymbol{\mu}^+ $.
	\item[$ \boldsymbol{\gamma}_{\min/\max} $] Shadow prices for generation constraints.
	\item[$ \sigma $] Shadow price for the linearized loss constraint.
	\item[$ l $] Total network loss.
	\item[$ l_0 $] Intercept for the linearized loss constraint.
	\item[$ LF $] Linearized factors for network loss. 
	\item[$ \boldsymbol{d} $] Loss distribution factor.
	\item[$ \boldsymbol{\pi},\boldsymbol{\pi}^{\mathcal{E}/\mathcal{C}/\mathcal{L}} $] The LMP vector and its marginal energy, congestion, and loss components, respectively. 
	\item[$ S,H $] Subspace and hyperplane, respectively.
	\item[$ M $] The number of time intervals.
	\item[$ k $] The number of ever-congested lines.
	\item[$ \boldsymbol{X} $] The congestion LMP data matrix. 
	\item[$ \boldsymbol{\chi} $] Vector of $ \boldsymbol{\mu} $ for the ever-congested $ k $ lines. 
	\item[$ \mathfrak{X} $] Congestion status. Boolean version of $ \boldsymbol{\chi} $.
	\item[$ \boldsymbol{b},\boldsymbol{B} $] Basis vector and matrix, respectively.
	\item[$ \boldsymbol{n} $] Norm vector of a hyperplane.
	\item[$ \boldsymbol{v} $] Example vector in the high-dimensional space.
	\item[$ \boldsymbol{A} $] Affinity or similarity matrix.
	\item[$ \gamma $] Eigenvalue in spectral clustering.
	\item[$ \xi $] Eigengap in spectral clustering.
	\item[$ c_i(\cdot) $] Cost function offered at node $ i $.
	\item[$ \mathds{1}(\cdot) $] Boolean function for a given condition.
	\item[$ ||\cdot||_{F,0,1} $] Frobenius norm, $ \ell_0 $ norm, and $ \ell_1 $ norm.
	\item[$ \kappa(\cdot) $] Elementwise cutoff kernel function.
	\item[$ span(\cdot) $] Subspace spanned by certain column vectors.
	\item[$ \boldsymbol{\Theta}(\cdot) $] Function that counts the number of zero entries of a given vector. 
\end{IEEEdescription}

\section{Introduction}
%
%
%
%

\IEEEpubidadjcol

\IEEEPARstart{T}{he locational} marginal prices (LMPs) are the foundation of modern electricity markets. They reflect the marginal operation and generation cost of the whole power system at each node~\cite{conejo2005locational,yang2017lmp}. The data of LMPs are calculated and publicized by an independent system operator (ISO) or regional transmission organization (RTO) according to the price bid \& offer collected in the day-ahead and real-time electricity market. The problem of security-constrained economic dispatch (SCED) is formulated and solved using linear programming or convex optimization tools. The system information, including network topology, transmission capacity, and line impedance, is included in the SCED and partially determines the distribution of LMP data. 

Generally, detailed market structure information (\textit{e.g.}, topology and line parameters) is private due to security concerns. Thus, the market participants may not have access to such information. Researchers aim to study the publicized LMP data with additional system information in data-driven approaches. In~\cite{geng2017learning}, Geng \textit{et al}. proposed a multi-class support vector machine (SVM)-based approach to fit the LMP-load coupling. The LMP in a certain power system was modeled as a function of loads at all nodes. The distribution of LMPs in the load space was analyzed and fitted precisely using an SVM. Kekatos \textit{et al}. aimed to recover network topology from LMP data in~\cite{kekatos2016online}. The low-rank feature of the LMP data and sparsity of the topology adjacency matrix were utilized to formulate a convex optimization problem. It has been proved that as long as enough LMP data are given, the connectivity of certain lines could be approximately recovered and tracked. In~\cite{birge2017inverse}, linear regression was used to recover the potential market structure based on the released LMP data and Lagrange multipliers of the inequality constraints in the SCED problem. However, in some electricity markets, the exact values of the Lagrange multipliers are not publicized.

Recent research on LMP~\cite{geng2017learning,mather2018} has revealed that if the network parameters and generation offers remain the same, the congestion component of the LMP data lies in a discrete space \textit{w.r.t.} nodal loads. In other words, for the nodal loads that belong to the same system pattern region (SPR), the network congestion status and the congestion LMP vector remains the same when the generation offers are constant. 
A simple proof using multiparametric linear programming (MLP) was provided in~\cite{geng2017learning}, where the SPR was defined as the critical region (CR) in the theory of MLP. A much more rigorous proof and discussion on the SPRs of LMP under both linear and quadratic cases were given in~\cite{mather2018}. 

Most previous works focus on learning the coupling between additional system information and LMP data. We want to utilize the aforementioned characteristics of the congestion LMP vectors and extract hidden messages from them without any additional system information (\textit{e.g.}, nodal load data or topology data, which are usually unavailable). 
The core work here is to identify the network congestion status/SPRs using only the publicized LMP data. 
If the SPR of the LMP is identified, the distribution of the congestion LMP vector in the high-dimensional Euclidean space is limited to certain subspaces, which could help in LMP forecasting~\cite{radovanovic2019holistic,zhou2011short}. Additionally, through congestion status identification, we can know the borderline conditions of the system parameters as well as load and generator quotations. These results can help market participants enhance their understanding of the market structure and thus help to determine their bidding strategy.

Some studies show the benefit of SPR/CR information in LMP forecasting and other applications. Zhou \textit{et al.} exploited the structural aspects underlying the power market operations using SPR to reduce the congestion forecast error~\cite{zhou2011short}. The convex hull of the zonal load was fitted in high-dimensional load space with the QuickHull algorithm. In~\cite{ji2016probabilistic}, the short-term probabilistic LMP forecasting problem was considered with load uncertainty from a system operator perspective. CRs were used to map the uncertainty from the load space to the LMP space.

Congestion status is also important for bidding in energy markets and financial transmission right (FTR) markets. For example, in~\cite{wang2017impact} the impact of local transmission congestion on energy storage arbitrage opportunities was discussed. A bilevel imperfect competition model was proposed and the results indicated that the network congestion had a large impact on the energy storage profits. In~\cite{fang2016strategic}, a strategic bidding method considering FTR and demand response was proposed for load serving entities. In~\cite{loprete2019virtual}, Lo Prete \textit{et al.} studied the usage of unprofitable virtual transactions aiming at altering the network congestion status and improving the FTR revenue based on an equilibrium model. The information of congestion scenarios is fundamental in these models. In~\cite{chen2019learning}, an inverse optimization-based approach for estimating rivals’ production cost functions in electricity markets was proposed, and a discussion on its extension was given to take into account the network congestion status.

A static generation offering behavior is usually assumed in the application of SPR/CR. 
Actually, the congestion LMP vector changes within the same subspace when the marginal generation offer fluctuates~\cite{cheverez2009admissible}. Thus, the discrete space that the LMP vector lies in is turned into a continuous space, and we cannot easily identify the congestion status from the LMP. In this paper, we propose an unsupervised method to cluster the LMP data into different subspaces in hierarchical ways with some efficient machine learning techniques. 
The efficacy of the method is proved by numerical experiments on the IEEE 30-bus system, IEEE 118-bus system, Illinois 200-bus system~\cite{birchfield2017grid}, and Southwest Power Pool (SPP)~\cite{spp} with real market data.
The contribution of this paper is threefold. 

\begin{enumerate}
	\item Problem modeling: We study the problem of identifying the congestion status of the LMP in a data-driven approach. The problem is modeled as recovering basis vectors for the congestion LMP data in high-dimensional Euclidean space.
	\item Highly efficient solution framework: By applying efficient machine learning techniques, including principal component analysis, spectral clustering, and dual principal component pursuit in bottom-up and top-down hierarchical searching, the hidden subspaces in LMP data are identified one by one. The bottom-up algorithm can directly find the basis vectors that span the subspaces when there is no basis gap. The top-down algorithm can find the potential subspaces by detecting the norm vectors and addressing the basis gap.
	\item Comprehensive experiments: Four numerical cases, including one with real market data, are presented to give a visualized procedure for the hierarchical searching methods. The cases verify the high accuracy and low time consumption of the proposed method.
\end{enumerate}

This paper is structured as follows. Section~\ref{sec:model} describes the widely adopted LMP models and analyzes the distribution of congestion LMP vectors. Section~\ref{sec:method} gives the proposed methodology and detailed algorithms. Three numerical cases are conducted in Section~\ref{sec:numerical}. Finally, Section~\ref{sec:conclusion} discusses the application of the proposed method and draws the conclusion.

\section{Model and Distribution of LMP}
\label{sec:model}
The DC optimal power flow (DC-OPF) model has been widely adopted in SCED. The lossless DC-OPF model is formulated as follows:
\begin{subequations}
	\label{equ:lossless}
	\begin{align}
	&\min_{\boldsymbol{P}_G} \sum_{i\in\mathcal{N}} c_i(P_{Gi}) & & \\
	\text{s.t. } &\boldsymbol{1}^\top \boldsymbol{P}_G = \boldsymbol{1}^\top \boldsymbol{P}_D &:&\lambda \label{equ:1-energy} \\
	&|\boldsymbol{T}(\boldsymbol{P}_G-\boldsymbol{P}_D)| \leq \boldsymbol{f}^{\max}  &:&\boldsymbol{\mu}^+,\boldsymbol{\mu}^- \label{equ:1-congest} \\
	&\boldsymbol{P}^{\min} \leq \boldsymbol{P}_G \leq \boldsymbol{P}^{\max} &:&\boldsymbol{\gamma}_{\min}, \boldsymbol{\gamma}_{\max}. \label{equ:1-gen}
	\end{align}
\end{subequations}
The LMP vector $ \boldsymbol{\pi} $ is the derivation of the Lagrangian function of (\ref{equ:lossless}) \textit{w.r.t.} $ \boldsymbol{P}_D $:
\begin{equation}
\label{equ:LMP1}
\boldsymbol{\pi} = \lambda\cdot \boldsymbol{1} + \boldsymbol{T}^\top \boldsymbol{\mu}.
\end{equation}
The LMP vector in~(\ref{equ:LMP1}) can be decomposed into two parts, the energy component $ \boldsymbol{\pi}^{\mathcal{E}} $ and the congestion component $ \boldsymbol{\pi}^{\mathcal{C}} $:
\begin{subequations}
	\begin{align}
	&\boldsymbol{\pi}^{\mathcal{E}} = \lambda_1 \boldsymbol{1}\,; \\
	&\boldsymbol{\pi}^{\mathcal{C}} = \boldsymbol{T}^\top \boldsymbol{\mu} .
	\end{align}
\end{subequations}

The lossy DC-OPF model has been a hot topic for years, and different ISOs have slightly different ways of handling network loss in LMP assignment~\cite{eldridge2017marginal}. Here, we use the widely adopted model from Litvinov \textit{et al.}~\cite{litvinov2004marginal}:

\begin{subequations}
	\label{equ:lossy}
	\begin{align}
	&\min_{\boldsymbol{P}_G} \sum_{i\in\mathcal{N}} c_i(P_{Gi}) & & \\
	\text{s.t. } &\boldsymbol{1}^\top \boldsymbol{P}_G = \boldsymbol{1}^\top \boldsymbol{P}_D + l &:&\lambda \label{equ:2-energy} \\
	&l = l_0 + LF^\top(\boldsymbol{P}_G-\boldsymbol{P}_D)&:&\sigma \\
	&|\boldsymbol{T}(\boldsymbol{P}_G-\boldsymbol{P}_D-l\cdot\boldsymbol{d})| \leq \boldsymbol{f}_{\max}  &:&\boldsymbol{\mu}^+,\boldsymbol{\mu}^- \label{equ:2-congest} \\
	&\boldsymbol{P}^{\min} \leq \boldsymbol{P}_G \leq \boldsymbol{P}^{\max} &:&\boldsymbol{\gamma}_{\min}, \boldsymbol{\gamma}_{\max}. \label{equ:2-gen}
	\end{align}
\end{subequations}
According to \cite{litvinov2004marginal}, the lossy LMP is derived as:
\begin{equation}
\label{equ:lmp-2}
\boldsymbol{\pi}= \sigma\cdot \boldsymbol{1} - \sigma\cdot LF + \boldsymbol{T}^\top \boldsymbol{\mu} - \boldsymbol{1}\cdot \boldsymbol{d}^\top \boldsymbol{T}^\top \boldsymbol{\mu},
\end{equation}
and an additional loss component $ \boldsymbol{\pi}^{\mathcal{L}} $ can be decomposed:
\begin{subequations}
	\label{equ:decomp}
	\begin{align}
	&\boldsymbol{\pi}^{\mathcal{E}} = \sigma\cdot \boldsymbol{1} \,;\\
	&\boldsymbol{\pi}^{\mathcal{C}} = \underbrace{\boldsymbol{T}^\top \boldsymbol{\mu}}_{\boldsymbol{\pi}_1} - \underbrace{\boldsymbol{1}\cdot \boldsymbol{d}^\top \boldsymbol{T}^\top \boldsymbol{\mu}}_{\boldsymbol{\pi_2}} \,; \label{subeq:congestion}\\
	&\boldsymbol{\pi}^{\mathcal{L}} = -\sigma\cdot LF .
	\end{align}
\end{subequations}

\begin{figure}[!t]
	\centering
	\includegraphics[width=0.43\textwidth]{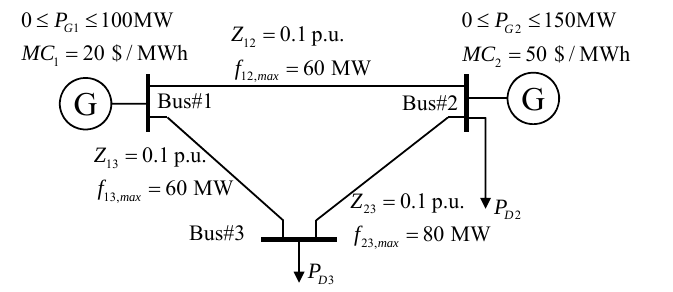}
	\caption{The 3-bus system in~\cite{geng2017learning}.}
	\label{fig:3-bus-system}
\end{figure}

Generally, ISOs release data of all three LMP components on their websites. In fact, only a very small proportion of lines in a transmission system can be congested during a short market period~\cite{radovanovic2019holistic}. In the California ISO, only two lines are typically congested~\cite{price2011reduced}. Thus, the vector of $ \boldsymbol{\pi}^{\mathcal{C}} $ always lies in the subspaces spanned by a few column vectors of matrix $ \boldsymbol{T}^\top $ (lossless) or $ (\boldsymbol{T}^\top - \boldsymbol{1}\cdot \boldsymbol{d}^\top \boldsymbol{T}^\top) $ (lossy). In lossy markets where the loss distribution vector $ \boldsymbol{d} $ is updated in a timely manner, the subspace feature of $ \boldsymbol{\pi}^{\mathcal{C}} $ might be unclear. In~(\ref{subeq:congestion}) the price consists of two parts, $ \boldsymbol{\pi}_1 $ spanned by constant vectors and $ \boldsymbol{\pi}_2 $ spanned by inconstant vectors. Note that $ \boldsymbol{1} $ is an all-ones vector and $ \boldsymbol{d}^\top \boldsymbol{T}^\top \boldsymbol{\mu} $ is a scalar, indicating that $ \boldsymbol{\pi}_2 $ is identical at every node. Thus, the effect of $ \boldsymbol{\pi}_2 $ can be eliminated by subtracting the value of $ \boldsymbol{\pi}^{\mathcal{C}} $ at a certain node from the $ \boldsymbol{\pi}^{\mathcal{C}} $ at every node.

\begin{figure}[!t]
	\centering
	\includegraphics[width=0.3\textwidth]{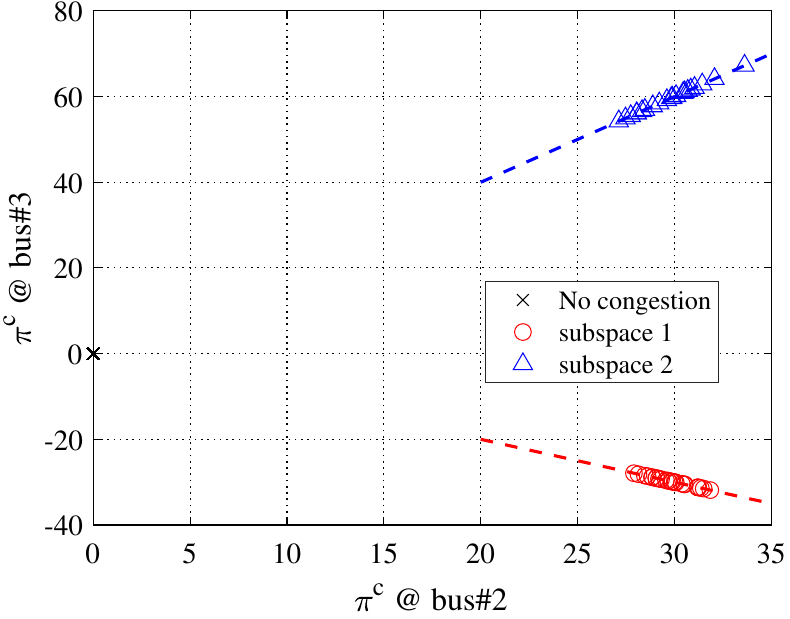}
	\caption{Two LMP patterns of the 3-bus system.}
	\label{fig:3node}
\end{figure}

Currently, some electricity markets are considering to adopt multi-interval SCED or look-ahead SCED~\cite{thatte2014analysis,hua2019pricing} in market operation. Ramping constraints and other complicated constraints would be included in the constraints of these SCED models. These constraints will impact the solution of dual variables $ \boldsymbol{\mu} $ and $ \sigma $. However, the formulation of (\ref{equ:lmp-2}) and
(\ref{equ:decomp}) will not be changed explicitly, which can be proved by applying Kuhn-Tucker conditions and similar steps in~\cite{litvinov2004marginal} for these SCED models (see Appendix for details). Our analysis and the proposed method in this paper are still applicable for look-ahead SCED models.

We use the 3-bus system in~\cite{geng2017learning} as an example (Fig.~\ref{fig:3-bus-system}) and assume that bus\#1 is the reference bus. A small Gaussian fluctuation is added to the marginal cost of the two generators. The loads at bus\#2 and bus\#3 are chosen from within the range of -150 to 300 MWh to generate different LMP vectors with the lossless model in~(\ref{equ:lossless}). Fig.~\ref{fig:3node} shows the distribution of the LMP vectors in Euclidean space. Two congestion statuses can be easily noticed when all the vectors with congestion fall in two linear subspaces.

\begin{figure}[!t]
	\centering
	\includegraphics[width=0.35\textwidth]{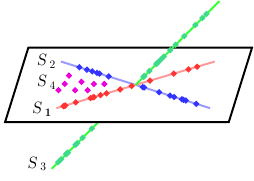}
	\caption{The overlapping subspaces: $ S_1 $, $ S_2 $, and $ S_3 $ have one congested line, while $ S_4 $ has two congested lines, corresponding to those of $ S_1 $ and $ S_2 $.}
	\label{fig:subspace}
\end{figure}

The LMP subspaces are more complex in a network with more nodes and transmission lines. It is hard to visualize these subspaces when the number of nodes is more than 4. However, we can refer to Fig.~\ref{fig:subspace} as a simplified illustration. The vector of $ \boldsymbol{\pi}^{\mathcal{C}} $ falls in subspace $ S_4 $ when both lines of $ S_1 $ and $ S_2 $ are congested together.

It is important to note that, when the system has only a few nodes, the LMP patterns (similar to those in Fig.~\ref{fig:3node}) can be easily identified using subspace clustering methods such as~\cite{vidal2005generalized,elhamifar2013sparse}. However, in real applications, the LMP patterns (similar to those in Fig.~\ref{fig:subspace} but with more complexity) are distributed in overlapping subspaces with outliers. The subspace $ S_4 $ in Fig.~\ref{fig:subspace} entirely covers $ S_1 $ and $ S_2 $, while most of the subspace clustering methods can handle only independent or disjoint subspaces~\cite{tierney2015segmentation,vidal2011subspace} and separate $ S_3 $ from $ S_4 $. Therefore, a specialized method for identifying overlapping LMP subspaces with outliers should be proposed.


\section{Methodology}
\label{sec:method}
The problem formulation, the basic framework of the proposed method, and detailed algorithms used are introduced in this section.

\subsection{Problem Formulation}
Assume that we have collected the time series of $ \boldsymbol{\pi}^{\mathcal{C}} $ at all nodes and denote it as $ \boldsymbol{X} $:
\begin{equation}
\boldsymbol{X} = [ \boldsymbol{\pi}^{\mathcal{C}}_t ]_{t\in \mathcal{T}}, \quad |\mathcal{T}| = M.
\end{equation}
Let $ j_1 $, $ j_2 $, $ \cdots $, $ j_k $ ($ k<<n_l $) denote the indexes of ever-congested lines during the time period $ \mathcal{T} $, and let $ T_1 $, $ T_2 $, $ \cdots $, $ T_{n_b} $ denote all the column vectors of matrix $ \boldsymbol{T}^\top $. As we mentioned in Section~\ref{sec:model}, $ \boldsymbol{X} $ can be expressed by the linear combination of $ T_{j_1} $, $ T_{j_2} $, $ \cdots $, $ T_{j_k} $:
\begin{equation} 
\label{equ:code}
\boldsymbol{X} = \big[\boldsymbol{x}_1, \boldsymbol{x}_2,\cdots, \boldsymbol{x}_M\big] = \big[T_{j_1}, T_{j_2},\cdots, T_{j_k}\big]\cdot \big[\boldsymbol{\chi}_t\big]_{t\in\mathcal{T}},
\end{equation}
where $ \boldsymbol{\chi}_t \in \mathds{R}^k $ is the linear coefficient determined by the dual variable of line congestion $ \boldsymbol{\mu} $ at time $ t $. Define $ \mathfrak{X}_t $ to show the nonzero entries of $ \boldsymbol{\chi}_t $:
\begin{equation}
\label{equ:boolcode}
\mathfrak{X}_t = \mathds{1}(\boldsymbol{\chi}_t \neq 0) \approx \mathds{1}(|\boldsymbol{\chi}_t| > \epsilon),
\end{equation}
where $ \epsilon $ is a small number. The value of $ \mathfrak{X}_t $ determines which congestion status the vector of $ \boldsymbol{\pi}^{\mathcal{C}}_t $ belongs to. If $ \mathfrak{X}_t $ is recovered from $ \boldsymbol{X} $, the congestion information of the electricity market at time $ t $ can be revealed. 
Theoretically there are $ 2^k $ possible values of $ \mathfrak{X}_t $, but usually the number of congestion statuses is much smaller in reality. In the numerical experiments of the IEEE 30-bus system in Section~\ref{sec:numerical}, there are only 9 appearing values of $ \mathfrak{X}_t $ with 4 possible congested lines, as shown in Table~\ref{tab:xt}.

\begin{table}[!t]
	\renewcommand{\arraystretch}{1.0}
	\caption{The appeared combinations of congested lines and values of $ \mathfrak{X}_t $ in the numerical test of IEEE case30.}
	\label{tab:xt}
	\centering
	\begin{tabular}{|c|c|c|}
		\hline
		\thead{Number of\\Congested lines} & Line No. & Value of $ \mathfrak{X}_t $  \\
		\hline
		4   &  \thead{\#(10,29,30,35)}   & \thead{$ (1,1,1,1)^\top $}   \\ \hline
		3   &  \thead{\#(10,30,35),\#(29,30,35)}   & \thead{$ (1,0,1,1)^\top, (0,1,1,1)^\top $}   \\ \hline
		2   &  \thead{\#(10,30),\#(10,35) \\ \#(29,35),\#(30,35) }   &  \thead{ $ (1,0,1,0)^\top $, $ (1,0,0,1)^\top $ \\ $(0,1,0,1)^\top $, $ (0,0,1,1)^\top $  }  \\ \hline
		1   &  \#30, \#35   &  \thead{ $(0,0,1,0)^\top $, $ (0,0,0,1)^\top $ } \\
		\hline
	\end{tabular}
\end{table}

\subsection{Basic Framework}

\begin{figure}[!t]
	\centering
	\includegraphics[width=0.35\textwidth]{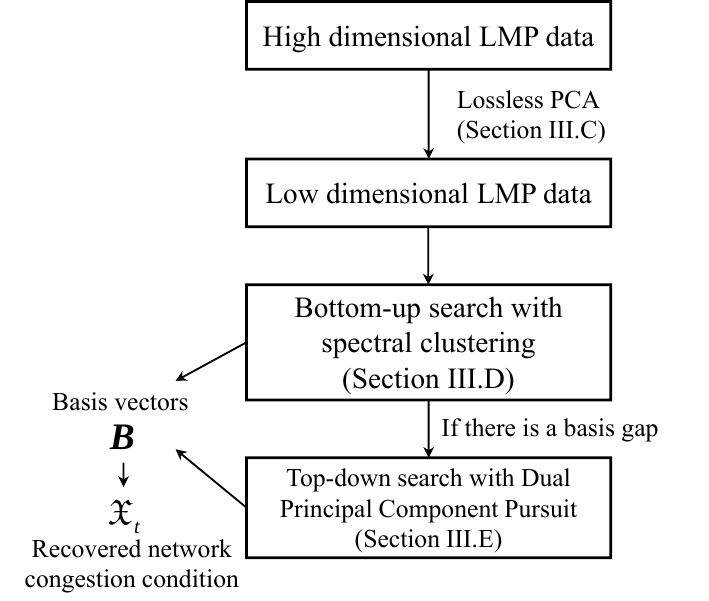}
	\caption{A simple framework of the proposed method.
	}
	\label{fig:framework}
\end{figure}

The basic target is to recover $ \mathfrak{X}_t $ from $ \boldsymbol{X} $ without any prior knowledge on $ \boldsymbol{T} $. 
Since all the $ M $ column vectors of $ \boldsymbol{X} $ lie in the subspace spanned by a common vector set $ \{T_{l_1}, T_{l_2},\cdots, T_{l_k}\} $, we need to identify the common basis vectors $ \boldsymbol{B} $ for $ \boldsymbol{X} $. First, the dimension of $ \boldsymbol{X} $ is reduced from $ n_b $ to $ k $, which reduces the computational burden for the following steps. Then a bottom-up approach is adopted to detect explicit basis vectors. If there are implicit basis vectors that result in a basis gap (introduced in the next subsection), a top-down approach is used to cover this situation. Once $ \boldsymbol{B} $ is recovered, $ \mathfrak{X}_t $ can be encoded as $ \mathds{1}(\boldsymbol{B}^\dagger \boldsymbol{x}_t \neq 0) $. Fig.~\ref{fig:framework} shows the basic framework of our method.

\subsection{Principal Component Analysis}

In a large-scale power system with a long-term price record, $ \boldsymbol{X} $ can be a high-dimensional matrix. Because the number of ever-congested lines $ k $ is usually small, $ \boldsymbol{X} $ is low rank, and its dimension can be reduced using principal component analysis (PCA). After PCA, $ k $ components can be used to represent the original data $ \boldsymbol{X} $ without any loss. By abuse of notation, we continue to use $ \boldsymbol{X} $ to denote the dimensionally reduced data matrix. The time complexity for PCA is $ O(n_b^2M+n_b^3) $. 

\subsection{Recursive Basis Search (Bottom-up)}
\label{subsec:bottom-up}
\begin{figure}[!t]
	\centering
	\includegraphics[width=0.30\textwidth]{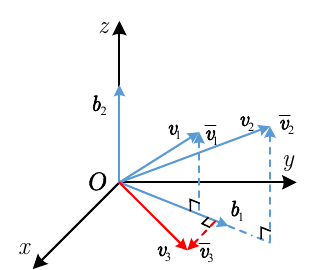}
	\caption{Bottom-up basis search using orthogonalization. $ \boldsymbol{b}_1 $ is the basis of the bottom subspace, and $ \boldsymbol{v}_1,\boldsymbol{v}_2 \in span(\boldsymbol{b}_1,\boldsymbol{b}_2)$ are in the upper subspace. $ \boldsymbol{v}_3 $ belongs to other subspaces. $ \bar{\boldsymbol{v}}_i (i=1,2,3) $ are the orthogonalized vectors \textit{w.r.t.} $ \boldsymbol{b}_1 $.
	}
	\label{fig:bottom-up}
\end{figure}

It is difficult to find all the overlapping subspaces that have different dimensions in $ \mathds{R}^k $ space with a large number of data points. However, it is much easier to identify 1-dimensional subspaces (\textit{i.e.}, straight lines passing through the origin) among these data points. If two vectors $ \boldsymbol{v}_1,\boldsymbol{v}_2 $ belong to the same 1-dimensional subspace, their cosine similarity should be $ 1 $:
\begin{equation}
\cos \angle (\boldsymbol{v}_1,\boldsymbol{v}_2) = \frac{\boldsymbol{v}_1^\top \boldsymbol{v}_2}{|\boldsymbol{v}_1||\boldsymbol{v}_2|}  =1.
\end{equation}
Thus, an affinity matrix $ \boldsymbol{A} $ can be constructed using cosine similarity. To filter out the similarity for those vectors belonging to different 1-dimensional subspaces, an elementwise cutoff kernel function
\begin{equation}
\kappa(x) = \begin{cases}
1, & (|x|>1-\varepsilon) \\
0, & \text{otherwise}
\end{cases}
\end{equation}
is used to obtain $ \boldsymbol{A}' = \kappa(\boldsymbol{A}) $, where $ \varepsilon $ is a small number. $ \boldsymbol{A}' $ is the sparse affinity matrix that contains only the affinity of vectors of the same 1-dimensional subspaces. Finding the subspaces in $ \boldsymbol{A}' $ is equivalent to finding the clusters according to $ \boldsymbol{A}' $. 
Then, a spectral clustering technique that maximizes the relative eigengap~\cite{sanchez2014hierarchical} can be used to identify the 1-dimensional subspaces automatically. Let $ \gamma_1,\cdots,\gamma_M $ denote the eigenvalues of $ \boldsymbol{A}' $ in ascending order. The relative eigengap is defined as:
\begin{equation}
\xi_i = \frac{\gamma_{i+1}-\gamma_i}{\gamma_i}, \quad i\geq 2. 
\end{equation}
A high value of $ \xi_i $ indicates that $ \boldsymbol{A}' $ admits a good decomposition into at least $ i $ clusters~\cite{sanchez2014hierarchical}. Thus, k-means with $ i $ as the number of partition sets is applied with the corresponding eigenvectors. 
The vector bases of these 1-dimensional subspaces can be calculated once the clusters are detected. 
Due to space limitation, the detail of the algorithm is not presented here. 

After we have found the data points that lie in the 1-dimensional subspaces, the next step is to search for the 2-dimensional subspaces. Take the vectors in Fig.~\ref{fig:bottom-up} as an example. Assume that $ \boldsymbol{v}_1, \boldsymbol{v}_2 $ belongs to the 2-dimensional subspace spanned by $ \boldsymbol{b}_1 $ and $ \boldsymbol{b}_2 $, while $ \boldsymbol{v}_3 $ lies in another subspace. If we have found vector $ \boldsymbol{b}_1 $ as a basis using spectral clustering, $ \boldsymbol{v}_1, \boldsymbol{v}_2, \boldsymbol{v}_3 $ can be orthogonalized into $ \bar{\boldsymbol{v}}_1,\bar{\boldsymbol{v}}_2,\bar{\boldsymbol{v}}_3 $ \textit{w.r.t.} $ \boldsymbol{b}_1 $ (i.e., projection onto $ span(\boldsymbol{b}_1)^\bot $). After orthogonalization, $ \bar{\boldsymbol{v}}_1 $ and $ \bar{\boldsymbol{v}}_2 $ lie in a 1-dimensional subspace, and the basis $ \boldsymbol{b}_2 $ can be easily calculated. This procedure can be repeated until all the subspaces and their bases are found. The pseudo code for the whole bottom-up searching procedure is shown in \textbf{Algorithm~\ref{algo:bottom-up}}.

The time complexity for calculating the affinity matrix $ \boldsymbol{A} $ is $ O(k^2M) $. For spectral clustering, full singular vector decomposition (SVD) is applied with time complexity of $ O(M^3) $. Then k-means clustering is used to group the vectors in the latent space.

Recall the structure in Table~\ref{tab:xt}. Our recursive basis search process is to determine the overlapping subspaces from the lowest one-dimensional space to the highest dimensional space (bottom-up). 
This method is powerful when dealing with structures with continuously increasing bases and will do well in situations where only a few lines are possibly congested. However, a problem arises when there is a \textbf{basis gap} after the orthogonalization, \textit{i.e.}, all the data points belong to subspaces whose dimension is higher than 1, as shown in Fig.~\ref{fig:top-down}. In this case, no obvious lines can be detected, and a top-down search method can be used to bridge the basis gap.


\begin{algorithm}[ht]
	\caption{Bottom-up subspace search}
	\label{algo:bottom-up}
	\begin{algorithmic}[1]
		\Procedure{Bottom-up}{$\boldsymbol{X}$, $ \varepsilon $}\Comment{$ \boldsymbol{X} $ has been dimensionally-reduced with PCA.}
		\\ \textit{Initialization} : $ \mathcal{B} \leftarrow \emptyset $;
		\Repeat 
		\State normalize $ \boldsymbol{X} $;
		\State $ \boldsymbol{A}\leftarrow |\boldsymbol{X}^\top \boldsymbol{X}| $; $ \boldsymbol{A}' \leftarrow \kappa(\boldsymbol{A},\varepsilon) $;
		\State $ \boldsymbol{X}_1,\cdots,\boldsymbol{X}_K $, and $ K \leftarrow $ \textsc{SpectralClustering}$ (\boldsymbol{X},\boldsymbol{A}') $;
		\For{$ i\gets 1:K $}
		\If{$ Rank(\boldsymbol{X}_i)==1 $}
		\State	$ \mathcal{B} \gets \mathcal{B}\cup \hat{\boldsymbol{X}}_i $
		\EndIf
		\EndFor
		\State $ \boldsymbol{X} \gets\text{Projection}(\boldsymbol{X},\mathcal{B}^\bot) $ \Comment{Dimension Reduction}
		\Until{$ K=1 $}
		\State \Return $ \mathcal{B} $
		\EndProcedure
	\end{algorithmic}
\end{algorithm}

\subsection{Hyperplane Detection (Top-down)}

\begin{figure}[!t]
	\centering
	\includegraphics[width=0.3\textwidth]{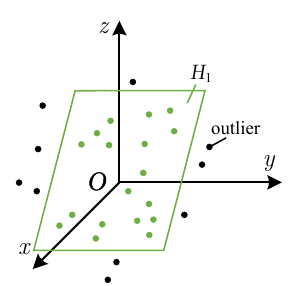}
	\caption{An example of the basis gap in $ \mathds{R}^3 $. $ H_1 $ is a 2-dimensional subspace where inliers are distributed. The outliers are in 3-dimensional space.
	}
	\label{fig:top-down}
\end{figure}

Assume that after the recursive basis search method in Section~\ref{subsec:bottom-up}, the distribution of data points is as in Fig.~\ref{fig:top-down}. The task of the top-down search is to find the basis vectors or the norm vector of $ H_1 $ among outliers with full rank. The problem of top-down searching can be formulated as follows:
\begin{equation}
\label{equ:DPCP}
\min_{\boldsymbol{n}} ||\boldsymbol{X}^\top \boldsymbol{n}||_0  \quad s.t.\quad ||\boldsymbol{n}||_2 = 1, 
\end{equation}
which aims to find the norm vector $ \boldsymbol{n} $ having as many points as possible contained in $ H_1 $. The $ \ell_0 $-norm in the objective function and the quadratic constraint in~(\ref{equ:DPCP}) make the problem nonconvex. Using any algorithm, there are three cases of possible solutions $ \boldsymbol{n} $ and $ H_1 $ for (\ref{equ:DPCP}):
\begin{enumerate}
	\item Global optimum. $ H_1 $ contains as many as possible data points.
	\item Non-trivial local optimum. $ H_1 $ contains some data points exceeding a certain threshold. 
	\item Trivial solution. $ H_1 $ contains only a few data points.
\end{enumerate}
For a global optimum or non-trivial local optimum, the result is acceptable because we have find a hyperplane that contains some data points and forms a congestion status.
Thus, we do not have to obtain the global optimum for~(\ref{equ:DPCP}) since a local optimal vector $ \tilde{\boldsymbol{n}} $ could also reveal a satisfactory subspace but containing a bit fewer points in it. 
The problem of~(\ref{equ:DPCP}) in high-dimensional space can be suitably addressed by a recently developed method in machine learning named dual principal component pursuit (DPCP)~\cite{tsakiris2015dual}. The brief procedure of DPCP is introduced here.

First, the $ \ell_0 $-norm of the objective function can be relaxed as the $ \ell_1 $-norm. However, the constraint of~(\ref{equ:DPCP}) is still nonconvex: 
\begin{equation}
\label{equ:DPCP2}
\min_{\boldsymbol{b}} ||\boldsymbol{X}^\top \boldsymbol{n}||_1  \quad s.t.\quad ||\boldsymbol{n}||_2 = 1. 
\end{equation}
To solve~(\ref{equ:DPCP2}), the constraint is further relaxed by a sequence of linear programs:
\begin{equation}
\label{equ:DPCP3}
\boldsymbol{n}_{i+1} := \underset{\boldsymbol{m}^\top \hat{\boldsymbol{n}}_i=1}{\mathrm{argmin}} ||\boldsymbol{X}^\top \boldsymbol{m}||_1, \quad i\in \mathds{N}, 
\end{equation}
where $ \hat{\cdot} $ indicates normalization to unit $ \ell_2 $-norm, and $ \boldsymbol{n}_0 $ is initialized to minimize $ ||\boldsymbol{X}^\top\boldsymbol{n}||_2 $ using SVD. The outliers and inliers can be separated according to their inner product with the converged normal vector $ \boldsymbol{n} $. In cases where the proportion of outliers is low, $ \{\boldsymbol{n}_i\} $ can converge to the global optimum of~(\ref{equ:DPCP2}).

Please note that there are several methods aiming at solving~(\ref{equ:DPCP2}) besides DPCP~\cite{tsakiris2015dual}, for example, Exact Recovery of Sparsely-Used Dictionaries (EP-SpUD)~\cite{spielman2012exact} and the algorithm based on Alternating Direction Method (ADM)~\cite{qu2014finding}. However, no algorithm is able to perfectly solve~(\ref{equ:DPCP2}) for any given $ \boldsymbol{X} $. DPCP is adopted in this paper because it performs well even in the challenging regime of large outlier ratios and high subspace relative dimensions~\cite{tsakiris2015dual,zhu2018dual}.

Assume that the dimension of the full-rank space is $ k $ and that the rank of $ H_1 $ is $ k-1 $. Let $ q $ denote the proportion of inliers in the total $ M $ points. It has been proved in~\cite{zhu2018dual} that DPCP can tolerate approximately $ O(\frac{1}{(k-1)k\log{k}^2}q^2M^2) $ outliers. However, if the number of inliers is relatively small, the algorithm cannot guarantee a positive result. 
When there are no positive results from DPCP, an alternative algorithm based on random sampling (RS) can be used to reveal the potential hyperplane. Randomly choose $ k-1 $ column vectors $ \boldsymbol{x}_{r_1},\cdots, \boldsymbol{x}_{r_{k-1}} $ from $ \boldsymbol{X} $ and the norm vector $ \tilde{\boldsymbol{n}} $ for $ span(\boldsymbol{x}_{r_1},\cdots, \boldsymbol{x}_{r_{k-1}}) $ can be constructed using Gaussian elimination or other algebraic methods. Then, check the following condition:
\begin{equation}
\label{equ:assert}
\boldsymbol{\Theta}(\boldsymbol{X}^\top \tilde{\boldsymbol{n}}) > p M, 
\end{equation}
where $ p $ is a predefined threshold. If~(\ref{equ:assert}) is satisfied, the hyperplane defined by $ \tilde{\boldsymbol{n}} $ contains at least $ p M $ points. Otherwise, the random sampling procedure is repeated until the maximum number of sampling $ N $ is reached. 

The probability of finding the expected norm vector in a single procedure is:
\begin{equation}
\frac{{qM\choose h-1}}{{M\choose k-1}} \approx q^{k-1}, \quad k-1 \ll qM. 
\end{equation}
The probability that the norm vector is not found within $ N $ procedures is
\begin{equation}
Prob = \Big(1-\frac{{qM\choose k-1}}{{M\choose k-1}} \Big)^N \approx (1-q^{k-1})^N. 
\end{equation}
If $ N $ is large enough, an upper bound for $ Prob $ can be guaranteed as $ Prob $ decreases exponentially \textit{w.r.t.} $ N $. The pseudo code for one iteration of the  top-down subspace identification method is given in \textbf{Algorithm~\ref{algo:top-down}}.

Since~(\ref{equ:DPCP3}) is a convex problem, it can be solved by efficient commercial solvers. The time complexity for RS is $ O(Nk^3)=O(\frac{\log\varepsilon}{\log{(1-p^{k-1})}}k^3)\approx O(-\log{\varepsilon}\cdot (\frac{1}{p})^{k-1}k^3) $ for keeping the failing probability $ Prob < \varepsilon $.

\begin{algorithm}[ht]
	\caption{Top-down subspace search}
	\label{algo:top-down}
	\begin{algorithmic}[1]
		\Procedure{Top-down}{$\boldsymbol{X}$,$ N $}
		\\ \textit{Initialization} : $ \tilde{\boldsymbol{n}} \leftarrow \emptyset $;
		\State $ \tilde{\boldsymbol{n}}\gets $DPCP($ \boldsymbol{X} $);
		\If{(\ref{equ:assert}) holds}
		\State \Return $ \tilde{\boldsymbol{n}} $
		\Else
		\For{$ i\gets 1:N $}
		\State $ \tilde{\boldsymbol{n}}\gets $RS($ \boldsymbol{X} $);
		\If{(\ref{equ:assert}) holds}
		\State \Return $ \tilde{\boldsymbol{n}} $
		\EndIf
		\EndFor
		\EndIf
		\EndProcedure
	\end{algorithmic}
\end{algorithm}

Theoretically, the top-down subspace search is applicable regardless of basis gap. However, using the bottom-up algorithm is better in the first place. As illustrated in Fig.~\ref{fig:case2} later, the top-down searching considers the problem in a tree structure that would have $ 2^k $ nodes in worst conditions. Thus, Algorithm~\ref{algo:top-down} is inefficient when $ k $ is large. If Algorithm~\ref{algo:bottom-up} is used first, part of the basis vectors might be obtained, and the dimension of data can be reduced to improve the overall efficiency.

\subsection{Short Summary}

\begin{figure}[!t]
	\centering
	\includegraphics[width=0.48\textwidth]{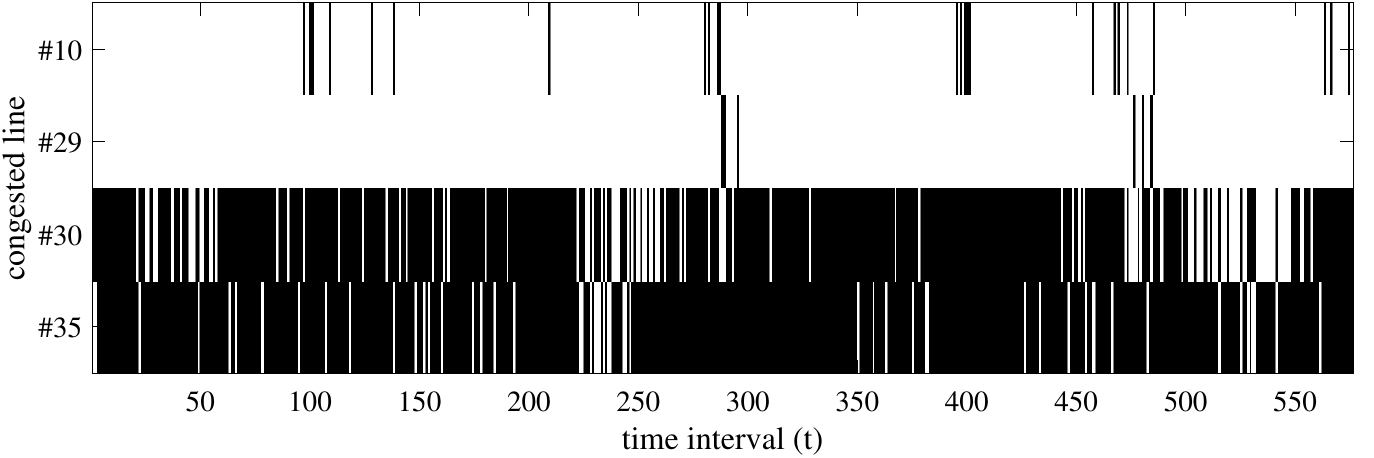}
	\caption{The 576 values of $ \mathfrak{X}_t $ in the numerical test of IEEE case30. Dark blocks indicate congestion (or $ 1 $-entries in $ \mathfrak{X}_t $), while light blocks indicate non-congestion (or $ 0 $-entries in $ \mathfrak{X}_t $).}
	\label{fig:xt_all}
\end{figure}

In summary, the whole process of our unsupervised congestion status identification method has the following steps: 
\begin{enumerate}
	\item Data preprocessing: remove the non-congestion vectors from matrix $ \boldsymbol{X} $. Use PCA for dimension reduction. 
	\item Run procedure \textsc{Bottom-up}$ (\boldsymbol{X},\varepsilon) $ in \textbf{Algorithm~\ref{algo:bottom-up}}, and obtain the basis set $ \mathcal{B} $. 
	\item Project $ \boldsymbol{X} $ onto $ \mathcal{B}^\bot $, and remove the all-zero columns from $ \boldsymbol{X} $. Let $ \boldsymbol{X}^{(2)} $ denote the result. Usually the process terminates here, because $ \boldsymbol{X}^{(2)} $ is empty.
	\item If $ \boldsymbol{X}^{(2)} $ is not empty, procedure \textsc{Top-down}$ (\boldsymbol{X}^{(2)},N) $ from \textbf{Algorithm~\ref{algo:top-down}} can be used iteratively to find the subspaces and their normal vectors. 
	\item Construct the basis matrix $ \boldsymbol{B} $ from the basis set $ \mathcal{B} $ of step 2 and the normal vectors of step 4. 
	If $ Rank(\boldsymbol{B}) $ equals $ Rank(\boldsymbol{X}) $, then we have all the basis to represent $ \boldsymbol{X} $. The dual variables $ \boldsymbol{\chi}_t $ can be recovered by:
	\begin{equation}
	\hat{\boldsymbol{\chi}}_t = \boldsymbol{B}^\dagger \boldsymbol{X}
	\end{equation}
	where $ \dagger $ means the Moore-Penrose pseudoinverse. And $ \mathfrak{X}_t $ can be recovered using (\ref{equ:boolcode}).
\end{enumerate}


\section{Numerical Experiments}
\label{sec:numerical}

In this section we show four cases based on the IEEE 30-bus system, the IEEE 118-bus system, the Illinois 200-bus system, and the SPP market, to demonstrate the effectiveness of the proposed method. All numerical experiments are conducted on an Intel Core i7-7500U@2.90 GHz laptop with \textsc{Matlab} R2016a. The market is cleared every 5 minutes. In each simulated case, 576 vectors of $ \boldsymbol{\pi}^{\mathcal{C}}_t $ with congestion are collected.

To evaluate the identification accuracy, assume that $ \boldsymbol{B} $ is the recovered basis matrix. We can recover $ \mathfrak{X}_t $ from $ \boldsymbol{B}^\dagger \boldsymbol{X} $. Let $ \mathfrak{C}_t $ denote the recovered code $ \mathds{1}(\boldsymbol{B}^\dagger \boldsymbol{x}_t) $, and the following miscode rate is used to evaluate the result:
\begin{equation}
misrate = 1 - \sum_{t\in \mathcal{T}} \frac{\boldsymbol{\Theta}(\mathfrak{X}_t-\mathfrak{C}_t)}{k\times M}. 
\end{equation}
For each row of $ \mathfrak{X}_t $, which corresponds to the congestion status of a single transmission line, the $ misrate $ can be calculated to show the identification accuracy. 

\subsection{Case 1: IEEE 30-bus system}
This case was conducted on the IEEE 30-bus system with 6 generations at buses \#1, \#2, \#13, \#22, \#23, and \#27. The block offers of the generations are set the same as in~\cite{kekatos2016online}. The load data at each bus are generated from a Kaggle open source dataset~\cite{kaggle2012} in January 2008. All the block offers and loads are added with Gaussian fluctuations of 3\% relative standard deviation. The corresponding values of $ \mathfrak{X}_t $ are shown in Fig.~\ref{fig:xt_all}. 

\begin{figure}[!t]
	\centering
	\includegraphics[width=0.48\textwidth]{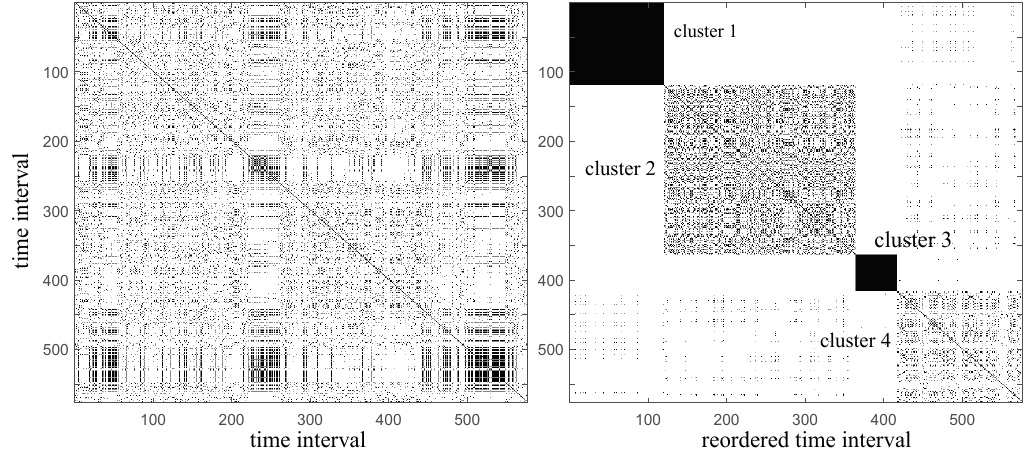}
	\caption{Bottom-up searching with spectral clustering. Left: the $ \boldsymbol{A} $ matrix with the cutoff kernel. Right: the spectral clustering result. Dark blocks indicate $ 1 $.}
	\label{fig:Amat}
\end{figure}

This case can be solved by bottom-up searching since there is no basis gap. After data preprocessing with PCA, the dimension of the data matrix $ \boldsymbol{X} $ is reduced to $ (4,576) $. Following the steps in \textbf{Algorithm~\ref{algo:bottom-up}}, we construct the $ \boldsymbol{A} $ matrix and apply the cutoff kernel $ \kappa $ with $ \varepsilon=0.005 $. Fig.~\ref{fig:Amat} shows the clustering results in the first iteration. Four clusters are detected, and clusters 1 and 3 are 1-dimensional subspaces. Thus, two basis vectors $ \boldsymbol{B}=[\boldsymbol{b}_1,\boldsymbol{b}_2] $ can be constructed from clusters 1 and 3. Then, $ \boldsymbol{X} $ is projected on the subspace $ \mathcal{B}^\bot = span(\boldsymbol{B})^\bot $, and the all-zero columns are removed to obtain $ \boldsymbol{X}^{(2)} $. 

\begin{figure}[!t]
	\centering
	\includegraphics[width=0.48\textwidth]{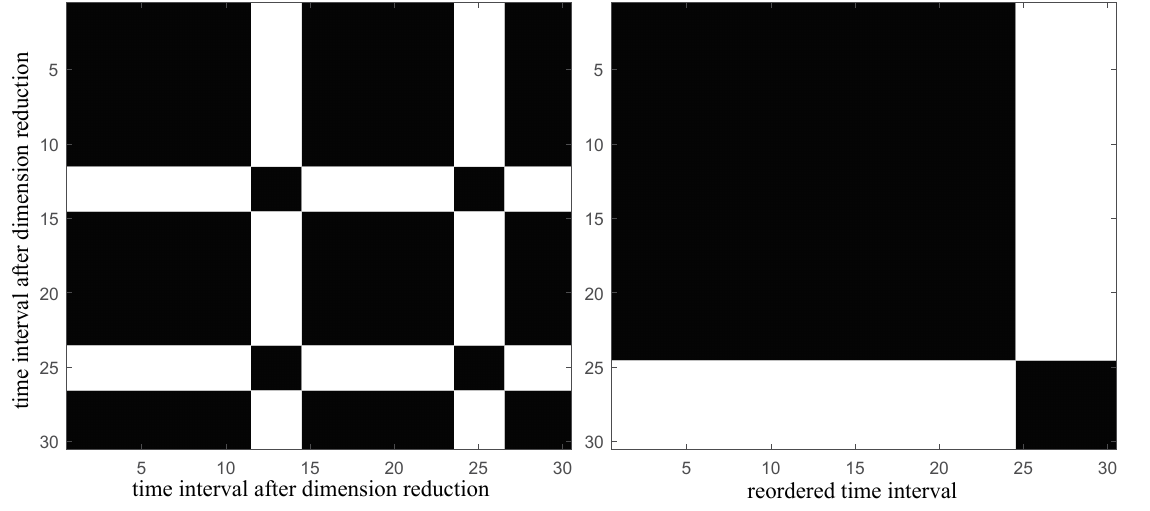}
	\caption{Bottom-up searching result in Round 2. Left: $ \boldsymbol{A} $ matrix updated from $ \boldsymbol{X}^{(2)} $. Right: $ \boldsymbol{A} $ matrix after spectral clustering and index reordering.}
	\label{fig:Amat2}
\end{figure}

In this case, $ \boldsymbol{X}^{(2)} $ still contains 30 points after one iteration. Thus, the bottom-up searching continues, and $ \boldsymbol{A} $ is updated in Round 2. Fig.~\ref{fig:Amat2} shows the clustering result; two clusters are detected in this round. Two additional basis vectors can be added to $ \boldsymbol{B} $. Now that $ Rank(\boldsymbol{B})=4 $, we can recover $ \mathfrak{X}_t $ from $ \boldsymbol{B}^\dagger \boldsymbol{X} $. As we can see from Table~\ref{table:misrate}, the miscode rate is very low for all the lines in this case, indicating a precise identification result. 

\subsection{Case 2: IEEE 118-bus system}

This case is intended to test the efficiency of the top-down searching algorithm. The regional load and load participation factors at different buses were set using the NREL extended case data of December 2012 provided in~\cite{pena2018extended}. To simplify the combinations of congested lines, we set transmission capacity for only 7 interregional lines. The block offers of the generators were set so that Region 1 had a higher generation cost than the other two regions, and 3\% Gaussian fluctuations were added to the load and block offers in the real-time market. 

\begin{table}[!t]
	\renewcommand{\arraystretch}{1.0}
	\caption{The appearing combinations of congested lines, values of $ \mathfrak{X}_t $, and their frequency in case 2.}
	\label{tab:xt118}
	\centering
	\begin{tabular}{|c|c|c|}
		\hline
		\thead{Number of\\Congested lines} & Line No. & Value of $ \mathfrak{X}_t $ and Frequency \\
		\hline
		4   &  \thead{\#(44,45,54,128)}   & \thead{$ (1,1,1,1)^\top $, 4.86\%}   \\ \hline
		3   &  \thead{\#(44,45,54)\\ \#(44,54,128)}   & \thead{$ (1,1,1,0)^\top$, 3.30\% \\ $(1,0,1,1)^\top $, 51.9\%}   \\ \hline
		2   &  \thead{\#(44,54)}  &  \thead{ $ (1,0,1,0)^\top $, 39.9\%}  \\ 
		\hline
	\end{tabular}
\end{table}

\begin{figure}[!t]
	\centering
	\includegraphics[width=0.45\textwidth]{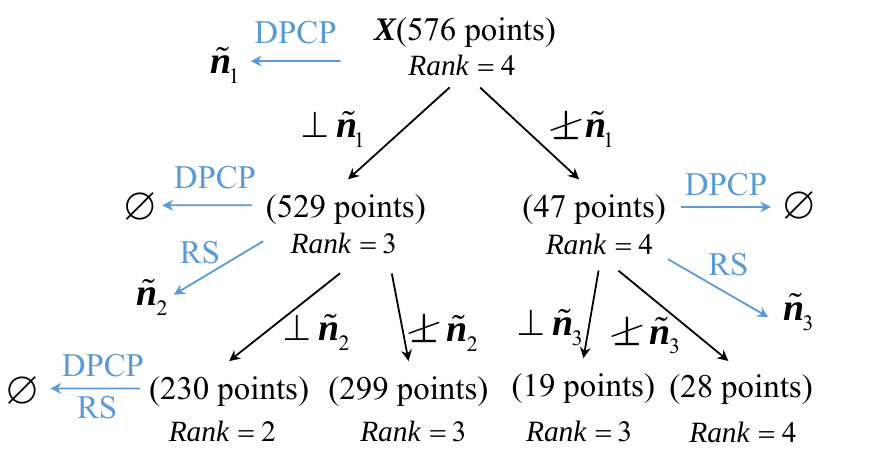}
	\caption{Iterative top-down searching process of case 2.}
	\label{fig:case2}
\end{figure}

Table~\ref{tab:xt118} shows the congestion combinations that appeared in this case. At least two lines are congested at any time, and the bottom-up searching algorithm cannot detect any rank-1 clusters. To address the basis gap in this case, the top-down searching algorithm is used. Fig.~\ref{fig:case2} shows the detailed searching process. First, \textbf{Algorithm~\ref{algo:top-down}} is applied to find the norm vector $ \tilde{\boldsymbol{n}}_1 $ for a hyperplane among the 576 points. The result of DPCP is positive, and the points are classified according to their orthogonality with $ \tilde{\boldsymbol{n}}_1 $. Here, 529 points lie in the detected hyperplane, and the searching continues with them. DPCP has a negative result in Round 2, and RS is used alternatively. $ \tilde{\boldsymbol{n}}_2 $ is detected by RS, and 230 points are orthogonal to $ \tilde{\boldsymbol{n}}_2 $. Additionally, RS is applied to the 47 outliers in Round 1 to reveal hidden subspaces. $ \tilde{\boldsymbol{n}}_3 $ is detected with 19 orthogonal points. In Round 3, both DPCP and RS give a negative result, which means we have reached the bottom of the subspaces. 

To construct the basis vectors $ \boldsymbol{B} $ from the hierarchical structure of Fig.~\ref{fig:case2}, we start from the 230 points in the bottom. $ \boldsymbol{b}_1,\boldsymbol{b}_2 $ are extracted from them, and $ \boldsymbol{b}_3 $ is extracted from the other 299 points with $ Rank=3 $. Finally, $ \boldsymbol{b}_4 $ is extracted from the 19 points. The miscode rate for each congested line and in total are shown in Table~\ref{table:misrate}.

\subsection{Case 3: Illinois 200-bus system}

To test the proposed method in a large power system, simulation of the electricity market is conducted on the Illinois 200-bus system~\cite{birchfield2017grid}. The load data are randomly generated based on the values given in the standard case in \textsc{Matpower}~\cite{zimmerman2011matpower}. The block offers of generators are set the same as two blocks of 65\% generator capacity with a price of 30 \$/MWh and 35 \$/MWh. Six of the 245 lines are ever-congested during the test time period. 

In this case, the basis gap also exists, and the top-down algorithm is used to identify all the subspaces. Similar to Case 2, DPCP and RS are used iteratively to conduct a top-down hierarchical classification of the whole dataset. After 5 rounds, 6 basis vectors are detected to form the basis matrix $ \boldsymbol{B} $. The miscode rates are shown in Table~\ref{table:misrate}.

\begin{table}[!t]
	\renewcommand{\arraystretch}{1.0}
	\caption{Miscode rates in the three numerical cases}
	\label{table:misrate}
	\centering
	\begin{tabular}{|c|c|c|c|c|}
		\hline
		$ misrate $ & \thead{IEEE\\30-bus system} & \thead{IEEE\\118-bus system} & \thead{Illinois\\200-bus system} & \thead{SPP\\market} \\
		\hline
		Total &  0.39\%  & 0 & 0 & 0.54\% \\
		1st Line & 0 & 0 & 0 & 0 \\
		2nd Line & 0 & 0 & 0 & 0.80\% \\
		3rd Line & 1.22\% & 0 & 0 & 1.61\% \\
		4th Line & 0.35\% & 0 & 0 & 0 \\
		5th Line & - & - & 0 & 0 \\
		6th Line & - & - & 0 & 0.80\% \\
		\hline
	\end{tabular}
\end{table}

\subsection{Case 4: Southwest Power Pool (SPP)}
\begin{figure}[!t]
	\centering
	\includegraphics[width=0.4\textwidth]{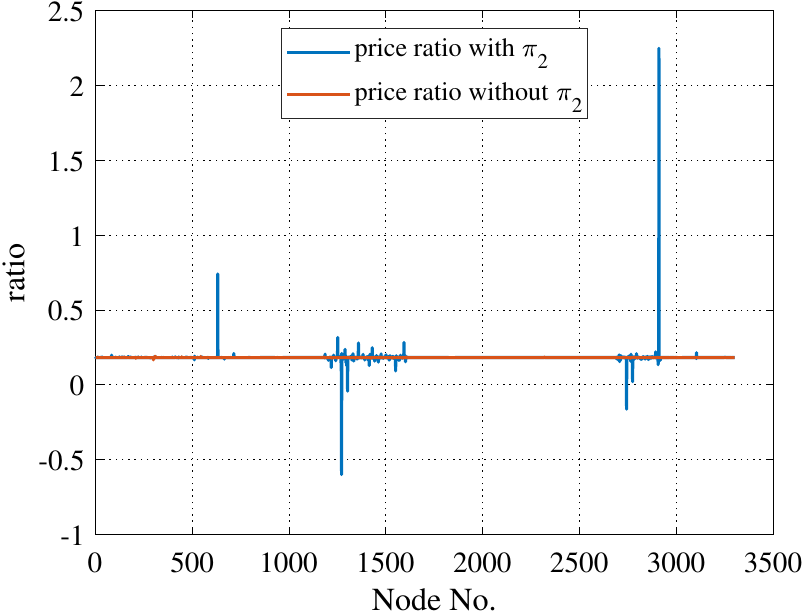}
	\caption{The nodal congestion price ratio of the 5th and 3rd operation interval in SPP on December 7, 2018.}
	\label{fig:ratio}
\end{figure}

We collected data from the real-time market of the SPP, which is an RTO serving 14 states in the USA. LMP data from 7\,408 electric nodes on December 7, 2018 are used. We merged the nodes that differed only in name and had identical LMPs all the time, leaving 3\,300 electric nodes. The SPP publishes the real-time $ \boldsymbol{\pi^{\mathcal{C}}} $ along with the binding constraints and their shadow prices $ \boldsymbol{\mu} $ on its website. Among the 288 time intervals, 249 intervals have one or more binding constraints. Six lines have congested once or more during the day. 
Since the SPP uses the lossy OPF model, the raw $ \boldsymbol{\pi^{\mathcal{C}}} $ data need to be preprocessed using the method mentioned in Section~\ref{sec:model} to eliminate the impact of $ \boldsymbol{\pi}_2 $. Fig.~\ref{fig:ratio} shows an example on the results of the elimination. The 3rd and 5th operation interval on December 7, 2018 belong to the same congestion status with only one line congested. The congestion price ratio are nearly the same for all the nodes after removing $ \boldsymbol{\pi}_2 $.

\begin{figure}[!t]
	\centering
	\includegraphics[width=0.48\textwidth]{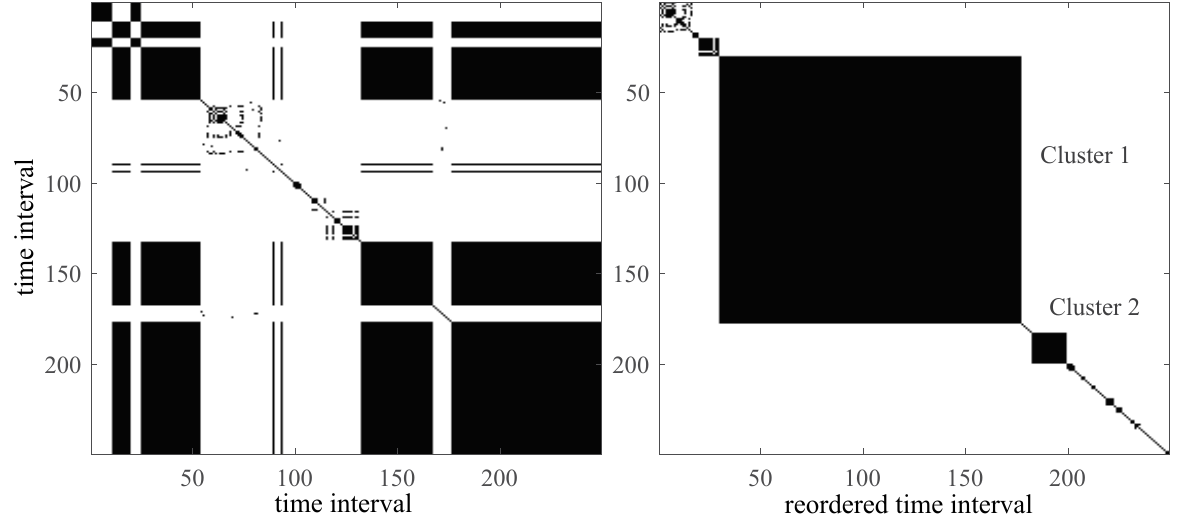}
	\caption{Bottom-up searching with SPP data in Round 1. Left: the original $ \boldsymbol{A} $ matrix. Right: $ \boldsymbol{A} $ matrix after spectral clustering and index reordering.}
	\label{fig:spp}
\end{figure}

\begin{figure}[!t]
	\centering
	\includegraphics[width=0.48\textwidth]{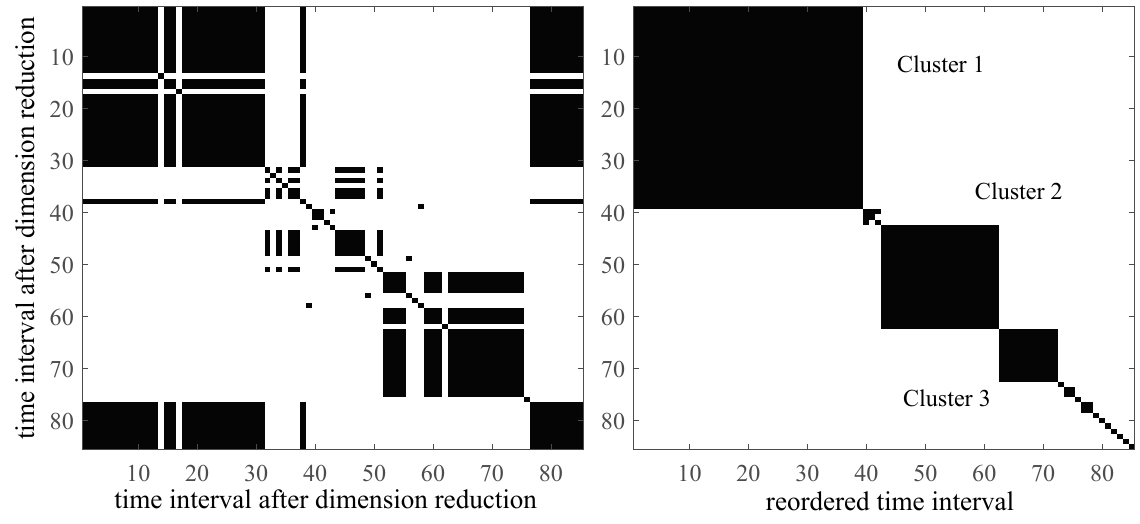}
	\caption{Bottom-up searching with SPP data in Round 2. Left: $ \boldsymbol{A} $ matrix updated from $ \boldsymbol{X}^{(2)} $. Right: $ \boldsymbol{A} $ matrix after spectral clustering and index reordering.}
	\label{fig:spp2}
\end{figure}

In this case, bottom-up searching is used. The clustering results in Round 1 are shown in Fig.~\ref{fig:spp}, and 2 rank-1 clusters are detected. The size of cluster 1 and cluster 2 is 147 and 17, respectively. Thus, we obtain 2 basis vectors after Round 1. There are still 85 time intervals after dimension reduction, and a new round of search starts for the remaining data. In Round 2, three basis vectors are detected, as shown in Fig.~\ref{fig:spp2}. The size of the three clusters are 39, 20, and 10, respectively. Finally, the last basis vector is found in Round 3. The detailed miscode rates are shown in Table~\ref{table:misrate}.

\subsection{Time consumption}

To test the time consumption of the proposed method in systems of different scales, we repeated the identification method in the three cases for 50 times. The average time consumption values are listed in Table~\ref{tab:time}. The entire procedure of the method costs only a few seconds or even less in some cases. 


\section{Discussion and Conclusion}
\label{sec:conclusion}
%
This section presents a discussion on the potential application of the proposed method and concludes the paper.

\subsection{Discussion on Potential Applications}

\begin{table}[!t]
	\renewcommand{\arraystretch}{1.0}
	\caption{Average time consumptions (Second) of the four cases.}
	\label{tab:time}
	\centering
	\begin{tabular}{|c|c|c|c|c|}
		\hline
		Step & \thead{IEEE\\30-bus system} & \thead{IEEE\\118-bus system} & \thead{Illinois\\200-bus system} &\thead{SPP\\market} \\
		\hline
		PCA & 0.002  & 0.006 & 0.013 & 0.08 \\
		Rd.1 & 0.0230 & 0.480 (DPCP) & 1.220 (DPCP) & 0.225 \\
		Rd.2 & 0.0150 & 1.105 (DPCP/RS) & 0.370 (DPCP) & 0.071 \\
		Rd.3 & - & - 	& 0.546 (DPCP) & 1E-4 \\
		Rd.4 & - & - 	& 0.366 (DPCP) & - \\
		Rd.5 & - & -	& 0.554 (DPCP/RS)& - \\ 
		\hline
	\end{tabular}
\end{table}

Essentially, the proposed method determines the network congestion status from public LMP data. The method has the
following applications for market participants.
\begin{itemize}
	\item Revealing the information of the congested status for each time period, especially for emerging markets where such information is not publicized, such as markets in China and Mexico. 
	\item LMP forecast calibration or reconciliation. The basis vectors obtained from the method are useful for LMP forecasting. The forecasted LMP data achieved from traditional LMP forecasting models (e.g., neural networks, SVM, etc.) do not naturally follow the distribution mentioned in Section~\ref{sec:model}. Thus, it would be necessary to reconcile the forecasting results according to the revealed basis vectors with quadratic programming~\cite{hyndman2016fast} or other related methods and get structurized forecasting results.
	\item Integrating with exogenous market variables. The congestion status is a function of nodal loads, generation bids, and transmission capacity. Such function can be fitted with exogenous market variables such as nodal loads~\cite{geng2017learning,zhou2011short} and generation mix~\cite{radovanovic2019holistic} and be further used for congestion forecasting, FTR valuation, and quantifying the market power of certain generators. 
\end{itemize}

\subsection{Conclusion}

This paper proposes a congestion status identification method based on unsupervised machine learning techniques with LMP data. The low-rank property of the congestion component of LMP is analyzed theoretically. Then the original problem is formulated as a subspace detection problem, which is solved by constructing the basis vector set and determining the coefficient of the basis vectors for each LMP data point. Two hierarchical basis vector searching algorithms are proposed to address the subspace detection problem. The bottom-up search can efficiently detect the basis vectors when there is no basis gap, and recover part of the power transfer distribution factor matrix of the power system. The top-down search can find the subspaces in cases with a basis gap. The results from four numerical cases show that the proposed method achieves nearly 100\% accuracy within only a few seconds. 

Future work includes integrating the identification method with LMP forecasting and market bidding strategy making. 

\appendix[Derivation of LMP for multi-interval SCED]
The general formulation of the multi-interval or look-ahead SCED considering linearized network loss and other more complicated constraints is given below:
\begin{subequations}
	\begin{align}
	& \min_{P_{G_i}[k]}\sum_{t=1}^{T}\sum_{i\in \mathcal{N}} c_i(P_{G_i}[t]) \\
	\lambda [t] : & \sum_{i\in \mathcal{N}} P_{G_i} [t] = \sum_{i\in\mathcal{N}} P_{D_i} [t] + l[t]\\
	\sigma [t]: & l[t] = l_0 [t] + LF[t]^\top (\boldsymbol{P}_G[t] - \boldsymbol{P}_D[t])\\ 
	\boldsymbol{\omega} [t] : & \Big| P_{G_i} [t] - P_{G_i} [t-1] \Big| \le R_i \Delta T,\, \forall i \label{subequ:ramping}\\
	\boldsymbol{\tau} [t] : & \boldsymbol{C}[t] \boldsymbol{P}_{G}[t] \le \boldsymbol{b}[t] \label{subequ:complicate}\\
	\boldsymbol{\gamma} [t] : & P^{\min}_{G_i} \le P_{G_i} [t] \le P^{\max}_{G_i},\, \forall i \\
	\boldsymbol{\mu} [t] : &  \boldsymbol{f}_{\min}\le \boldsymbol{T}(\boldsymbol{P}_G[t]-\boldsymbol{P}_D[t] - l[t]\cdot \boldsymbol{D}[t])  \le  \boldsymbol{f}_{\max}
	\end{align}
\end{subequations}
where $ t $ is the index for time intervals and $ R_i $ is the ramping capacity of generator at node $ i $. 
We use a general form of constraints as (\ref{subequ:complicate}) to denote other complicated constraints that are not inter-temporal.
The Lagrangian function of the aforementioned model is written as:
\begin{equation}
\begin{aligned}
\mathcal{L} & = \sum_{t=1}^{T}\sum_{i\in \mathcal{N}} c_i(P_{G_i}[t])\\
& - \sum_{t=1}^{T} \lambda [t] \Bigg[ \sum_{i\in \mathcal{N}} P_{G_i} [t] - \sum_{i\in\mathcal{N}} P_{D_i} [t] - l[t]\Bigg] \\
& - \sum_{t=1}^{T} \sigma [t] \Big[ l[t] - l_0 [t] - LF[t]^\top (\boldsymbol{P}_G[t] - \boldsymbol{P}_D[t])  \Big] \\
& + \sum_{t=1}^{T}\sum_{i\in \mathcal{N}} \Big[\omega_{i,\max}[t] (P_{G_i} [t] - P_{G_i} [t-1] - R_i \Delta T) \Big] \\
& + \sum_{t=1}^{T}\sum_{i\in \mathcal{N}} \Big[\omega_{i,\min}[t] (P_{G_i} [t-1] - P_{G_i} [t] - R_i \Delta T) \Big] \\
& + \sum_{t=1}^{T} \boldsymbol{\tau}_t \Big(\boldsymbol{A}[t] \boldsymbol{P}_G[t] - \boldsymbol{b}[t] \Big) \\ 
& + \sum_{t=1}^{T}\sum_{i\in \mathcal{N}} \Big[\gamma_{i,\max}[t] (P_{G_i}[t] - P_{G_i}^{\max}) \Big] \\
& + \sum_{t=1}^{T}\sum_{i\in \mathcal{N}} \Big[\gamma_{i,\min}[t] (P_{G_i}^{\min} - P_{G_i}[t]) \Big] \\ 
& + \sum_{t=1}^{T}\sum_{\ell =1}^{L} \Big[\mu_{\ell,\max}[t] (T_\ell (\boldsymbol{P}_G[t]-\boldsymbol{P}_D[t] - l[t]\cdot \boldsymbol{d}[t])  - f^{\max}_{\ell})  \Big] \\
& + \sum_{t=1}^{T}\sum_{\ell =1}^{L} \Big[\mu_{\ell,\min}[t] (f^{\min}_{\ell} - T_\ell (\boldsymbol{P}_G[t] - \boldsymbol{P}_D[t] - l[t]\cdot \boldsymbol{d}[t])  )  \Big] \\
\end{aligned}
\end{equation}
where $ \ell $ is the index for lines. According to the definition of LMP, we have:
\begin{equation}
\begin{aligned}
\pi_{i} [t] &= \frac{\partial  \mathcal{L} }{ \partial P_{d_i}[t] } = \lambda[t]  - \sigma [t] LF_{i}[t] \\
& - \sum_{\ell=1}^L \Big(\mu_{\ell,\max}[t] -\mu_{\ell,\min}[t]\Big)  T_{\ell,i}
\end{aligned}
\end{equation}
The vectorized form is:
\begin{equation}
\label{equ:vec_lmp1}
\boldsymbol{\pi} [t] = \lambda[t]\cdot \boldsymbol{1} - \sigma [t] LF[t] - \boldsymbol{T}^\top \Big(\boldsymbol{\mu}_{\max}[t] -\boldsymbol{\mu}_{\min}[t]\Big)
\end{equation}
From the first-order KKT condition:
\begin{equation}
\label{equ:loss}
\begin{gathered}
0 = \frac{\partial \mathcal{L}}{\partial l[t]}   = \lambda[t] - \sigma [t] - \sum_{\ell=1}^L \Big(\mu_{\ell,\max}[t]-\mu_{\ell,\min}[t] \Big) \cdot \\  T_\ell \boldsymbol{d} [t] 
 = \lambda[t] - \sigma [t] - \boldsymbol{d}[t]^\top \boldsymbol{T}^\top \Big(\boldsymbol{\mu}_{\max}[t] -\boldsymbol{\mu}_{\min}[t]\Big) \\
\end{gathered}
\end{equation}
Substituting (\ref{equ:loss}) into (\ref{equ:vec_lmp1}), we have:
\begin{equation}
\label{equ:pi}
\begin{aligned}
\boldsymbol{\pi}_i [t]  & =  \sigma [t] \boldsymbol{1} - \sigma [t] LF[t] - \boldsymbol{T}^\top \Big(\boldsymbol{\mu}_{\max}[t] -\boldsymbol{\mu}_{\min}[t]\Big)\\ &  + \boldsymbol{1} \cdot \boldsymbol{d}[t]^\top \boldsymbol{T}^\top \Big(\boldsymbol{\mu}_{\max}[t] - \boldsymbol{\mu}_{\min}[t]\Big) \\
\end{aligned}
\end{equation}
which does not differ from (\ref{equ:lmp-2}) explicitly.


%


\ifCLASSOPTIONcaptionsoff
  \newpage
\fi



%

\bibliographystyle{IEEEtran}
\bibliography{ref}

%
\begin{IEEEbiography}[{\includegraphics[width=1in,height=1.25in,clip,keepaspectratio]{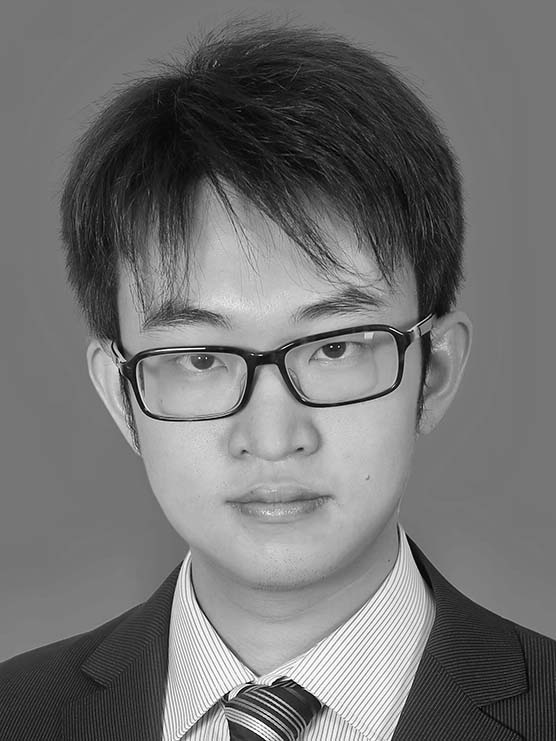}}]{Kedi Zheng} (S'17) received the B.S. degree in electrical engineering and automation from the Department of Electrical Engineering, Tsinghua University, Beijing, China, in 2017, where he is currently working toward the Ph.D. degree in electrical engineering. 
	
His research interests include application of big data analytics for electricity market.
\end{IEEEbiography}

\begin{IEEEbiography}[{\includegraphics[width=1in,height=1.25in,clip,keepaspectratio]{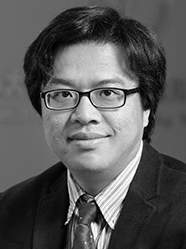}}]{Qixin Chen} (M'10–SM'15) received the Ph.D. degree in electrical engineering from the Department of Electrical Engineering, Tsinghua University,	Beijing, China, in 2010. 
	
He is currently an Associate Professor with Tsinghua University. His research interests include electricity markets, power system economics and optimization, low-carbon electricity, and power generation expansion planning.
\end{IEEEbiography}

\begin{IEEEbiography}[{\includegraphics[width=1in,height=1.25in,clip,keepaspectratio]{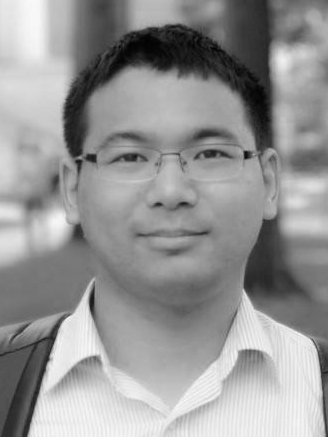}}]{Yi Wang} (S'14–M'19) received the B.S. degree from the Department of Electrical Engineering, Huazhong University of Science and Technology,	Wuhan, China, in 2014, and the Ph.D. degree from Tsinghua University, Beijing, China, in 2019. He was also a visiting student researcher with the 	University of Washington, Seattle, WA, USA from 2017 to 2018. 
	
He is currently a Post-Doctoral Researcher with ETH Z\"urich. His research interests include data analytics in smart grid and multiple energy systems.
\end{IEEEbiography}

\begin{IEEEbiography}[{\includegraphics[width=1in,height=1.25in,clip,keepaspectratio]{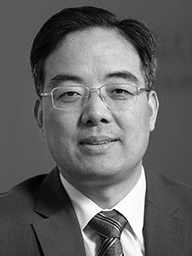}}]{Chongqing Kang} (M'01–SM'08–F'17) received the Ph.D. degree in electrical engineering from the Department of Electrical Engineering in Tsinghua University, Beijing, China, in 1997. 
	
He is currently a Professor with Tsinghua University. His research interests include power system planning, power system operation, renewable energy, low carbon electricity technology,	and load forecasting.
\end{IEEEbiography}

\begin{IEEEbiography}[{\includegraphics[width=1in,height=1.25in,clip,keepaspectratio]{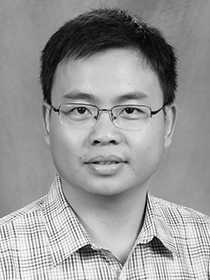}}]{Le Xie} (S'05–M'10–SM'16) received the B.E. degree in electrical engineering from Tsinghua University, Beijing, China, in 2004, the M.S. degree in engineering sciences from Harvard University, Cambridge, MA, USA, in 2005, and the Ph.D. degree from the Department of Electrical and Computer Engineering, Carnegie Mellon University, Pittsburgh, PA, USA, in 2009. 
	
He is currently a Professor with the Department of Electrical and Computer Engineering, Texas A\&M University, College Station, TX, USA. His research interests include modeling and control of large-scale complex systems, smart grids application with renewable energy resources, and electricity markets.
\end{IEEEbiography}





\vfill


\end{document}